\documentstyle[prd,aps]{revtex}
\begin{document}
\title{Different hierarchy of avalanches observed in Bak-Sneppen evolution model}
\author{W. Li$^{1,}$\thanks{E-mail: liw@iopp.ccnu.edu.cn} 
    and X. Cai$^{1,2,3,}$\thanks{E-mail: xcai@wuhan.cngb.com} \\ 
\footnotesize \sl 
$^{1}$Institute of Particle Physics, Hua-zhong Normal University, 
Wuhan 430079, China\\
$^{2}$INP, Universit\'e de Paris-Sud and CNRS-IN2P3, 
F-91406 Orsay Cedex, France\\
$^{3}$CERN, CH-1211, Geneva 23, Switzerland}

\maketitle
\vskip 0.5cm
\begin{abstract}
A new quantity, $\bar f$, denoting the average fitness of an ecosystem, 
is introduced in Bak-Sneppen model. Through this new quantity, a different 
hierarchy of avalanches, $\bar f_0$-avalanche, is observed in the evolution 
of Bak-Sneppen model. An exact gap equation, governing the self-organization 
of the model, is presented in terms of $\bar f$ . It is found that 
self-organized threshold $\bar f_c$ can be exactly obtained. Two basic 
exponents of the new avalanche, $\tau$, avalanche distribution, and $D$, 
avalanche dimension are given through simulations of one- and two-dimensional 
Bak-Sneppen models. It is suggested that $\bar f$ may be a good
quantity in determining the emergence of criticality. \\   

\noindent
PACS number(s): 87.10.+e, 05.40.+j, 64.60.Lx
\end{abstract}

\twocolumn

The term of avalanche may originate from the phenomena occurred in nature.
It is referred to as a sequential events which may cause devastating
catastrophe. The phenomena of avalanches are ubiquitous in nature. The
canonical example of avalanche is the mountain slide, during which
great mass of snow and ice at a high altitude slide down a mountain side,
often carrying with it thousands of tons of rock, and sometimes destroy
forests, houses, etc in its path [1]. Since avalanches occur everywhere,
from the ricepile, to the Himalayan sandpiles; from the
river network, to the earthquake, starquakes and even solar flares; from the
biology, to the economy, [2], etc, it is hence proposed [2] that avalanches 
may be the underlying mechanism of the formation of various geographical
structures and complex organisms, e.g., brains, etc. (It is now even
 proposed by Meng et al [3] that the formation of colorless gluon clusters
may be attributed to avalanches intrigued by emission or absorption of
gluons.) From this point of view,
avalanches can be viewed as the immediate results of complex systems, and 
hence can be used as the theoretical justification for catastrophism. This
is because if the real world is complex then the catastrophes are inevitable
and unavoidable. 

Plenty of patterns provided by nature exhibit coherent macroscopic
structures developed at various scales and do not exhibit elementary
interconnections. They immediately suggest seeking a compact description of
the spatio-temporal dynamics based on the relationship among macroscopic
elements rather than lingering on their inner structure [4]. That is, one
needs to condense information when dealing with complex systems. Maybe only
this way is efficient and turns out successful.

As known, avalanche is a kind of macroscopic phenomenon driven by local
interactions. The size of avalanche, spatial and temporal as well, may be
sensitive to the initial configuration, or more generally, the detailed
dynamics of the system. However, the distribution of avalanches, i.e., 
Gutenberg-Ritcher law [5], or equivalently, power-law, does not depend on such
kind of details due to the universality of complexity. Hence, in this sense
avalanche study may be an appropriate tools in studying various complex
phenomena. On the other hand, observation of a great variety of patterns,
such as self-similar, fractal behavior in nature [6,7,8,9], $1/f$
noise in quasar [10], river flow [11] and brain activity [12], and many
natural and social phenomena, including earthquakes, economic activity and
biological evolution suggests that these phenomena are signatures of
spatiotemporal complexity and can be related via scaling relations to the
fractal properties of the avalanches [13]. This suggestion means that 
the occurrence of 
these general, empirical phenomena may be attributed to the same underlying
avalanche dynamics. Thus, one can see that study of avalanche is crucial in
investigating the critical features of complex systems. It can be even
inferred that avalanche dynamics does provide much useful information for us
to understand the general features of the ubiquitous complexity around us.

Despite the fact that avalanche may provide insight into complexity, the
definition of which can be vastly different for various systems, and the
same sorts of systems, even the same system. Let us recall some definitions
of avalanche given before. In sandpile model [2], an avalanche is intrigued
by adding a grain or several grains of sand into the system at some time 
and causing the topple of some sites, which may later on cause some other
sites to topple. The avalanche is considered over when the heights of all
the sites are less than the critical value, say, 4. In Bak-Sneppen model [14],
several kinds of avalanche [13] are presented. For instance, $f_0$-avalanche, 
$G(s)$-avalanche, forward avalanche, backward avalanche, etc. 
Though these kinds of definitions of avalanche may show various
hierarchical structures they manifest the same underlying fractal feature of
the system, i.e., self-organized criticality (SOC). Relating all these kinds of
avalanches one can provide a general definition of the avalanche for Bak-Sneppen
model: An avalanche corresponds to sequential mutations below certain
threshold. One can see that this kind of definition can ensure the mutation
events within a single avalanche are casually and spatially connected. In
addition, with this definition there exists a hierarchy of avalanches, each
defined by their respective threshold. It is the hierarchical structure of
the avalanche,  which exhibits the fractal geometry of the system and 
implies complexity.                  

It can be inferred from the definition of avalanche that there always exists a
triggering event which initiates an avalanche and whose effect, that is,
causing the avalanche to spread later on within the system , will
disappear at the end of the avalanche. Observation of avalanche
through the triggering event, up to now, is based on the individual level,
despite that the avalanche is a macroscopic and global phenomenon of the
system studied, in the laboratory, and in nature as well. In sandpile
model, the triggering event is adding a grain or several grains of sand to
some sites and causing them to topple.
In Bak-Sneppen model, the corresponding
triggering event of an avalanche is mutation of the extremal species causing
the fitness [14] of the extremal site at the next time step less than a
certain threshold. In the above two models triggering
events are directly connected with the feature of individuals, e.g., the
height of the site in the former, or, the fitness of the extremal site
in the latter. It can be readily learned that the triggering events,
whether in the laboratory or in nature, are not directly related
to the global feature of the systems although avalanche can span the
whole systems. Generally speaking, the behavior of avalanches is observed
through the features of individuals, instead of those of the whole system. 
However, general features of the complex system may provide insight
into knowing the tendency of the evolution of the system. Specifically,
global features of a complex system may enable one to understand the critical
behavior of the system. It implies that some characteristic
quantities, representing the corresponding global features, can be employed in
describing complex systems. Furthermore, these
quantities ought to be related to avalanche dynamics, and hence can be used
to describe complexity emerged in a variety of complex systems. Apparently,
our aim is to search for or define such kind of quantities and 
expect to observe new types of avalanches based on these quantities. Indeed,
we obtain a new quantity which can be used to define a different hierarchy of
avalanches in Bak-Sneppen model. We suggest that this quantity may be used
as a criterion in determining the emergence of criticality. It will be shown
later that this new type of avalanche still exhibits spatio-temporal
complexity in another context.

Bak-Sneppen model [14] is a very simple evolution model of
biology. Despite the simplicity of the model itself, it can
exhibit the skeleton of species evolution---punctuated equilibrium.
In Bak-Sneppen model, each species is represented by a
single fitness. The fitness may represent population of a whole species or
living capability of the species [15]. It is a
vital quantity and the only one describing the model. No other additional 
quantities are employed in this oversimplified model. Thus, the fitness is
the most important feature of species, 
and absolutely the most important of the model. So, when
considering global feature of the ecosystem, one has to relate
it to that of individuals, i.e., fitness.

In this "toy" model (Bak-Sneppen model), random numbers, $f_i$, chosen from a flat
distribution, $p(f)$, are assigned independently to each species located on
a $d$-dimensional lattice of linear size $L$. At each time step, the extremal
site, i.e., the species with the smallest random number, and its
$2d$ nearest neighboring sites, are assigned $2d+1$
new random numbers also chosen from $p(f)$.
This updating process continues indefinitely. After a long transient process
the system reaches a statistically stationary state where the density of
random numbers in the system vanishes for $f < f_c$ and is uniform above
$f_c$  (the self-organized threshold).

Introduce a new quantity for Bak-Sneppen model. Define the
average fitness, denoted by $\bar f$, as,
\begin{equation}
\bar f=\frac {1} {L^d} \sum \limits_{i=1}^{L^d} f_i ,
\end{equation}
\noindent
where $f_i$ is the fitness of the $i$th species. Here, we refer to $\bar
f$ as the average fitness and a global quantity of the system.
$\bar f$ may represent average population or average living capability
of the whole ecosystem. Large $\bar f$, i.e., high average fitness, 
may imply the
total population of the system is immense or its average living capability
is great, and vice versa. Initial value of $\bar f$, denoted by $\bar f (0)$
, can be easily obtained. As known, at the beginning of the evolution 
$f_i$'s are   uniformly distributed between (0,1). 
Hence, for an infinite-size system, $\bar f (0)$ equals to 0.5. 
However, for a finite-size system $\bar f (0)$ fluctuates 
slightly due to the finite size effect. 
It should be pointed out that
$\bar f (0)$ does not reflect the correlation between species. As the
evolution goes on such correlation tends to 
be more distinctive. Denote $\bar f (s)$ the average fitness of
the system at time step $s$ in the evolution. In the $s$ limit, i.e.,
$s \gg L^d$, $\bar f (s)$ may partly reflect information about correlation.
As a global quantity, $\bar f (s)$ should include information concerning the
interaction between species. Hence, it is natural to expect that $\bar f$
may be a good quantity in describing the feature of the system as a whole.

Before introducing the different hierarchy of avalanches it is necessary and
worthwhile to investigate features of the new quantity, $\bar f (s)$. Firstly,
let us present some theoretical analysis. Recall the definition of $\bar f$
one can see that $\Delta \bar f (s)=\bar f (s+1)-\bar f (s)$ approaches zero
in the $L \rightarrow \infty$ limit. An observer can hardly perceive the
change in $\bar f (s)$ during such a short time period since it is 
vanishingly small.    
However, changes at each time step are accumulated to form a relatively
distinctive change after a long time, which is perceivable for the
observer. This long time period is required to be much larger 
than the system size, i.e.,
$s \gg L^d$. In other words, $\bar f(s+s_0)-\bar f(s_0)$ may only be
"noticed" when $s \gg L^d$ ($s_0$ denotes any initial time step). 
The variation of $\bar f (s)$ is small between two successive time steps,
which differs from that of fitness of extremal site. 
The latter can be very large, say, 1. It should also be expected that there
exists an increasing tendency of $\bar f (s)$ versus time $s$.
This is because at each time step the least fitness is eliminated from the
system so the general fitness of the whole system will tend to increase. And
due to the slow fluctuation of $\bar f (s)$ the increasing in $\bar f (s)$ 
behaves like a
stepwise, i.e., Devil's stepwise [2]. Hence, one may expect to observe such
behavior, i.e., punctuated equilibrium [14], of $\bar f (s)$ in the 
evolution of Bak-Sneppen model.

In order to show the feature of $\bar f(s)$ versus time $s$ we perform
simulations of Bak-Sneppen model. At each time step, in addition to the
updating of the extremal sites, we also track the signals $\bar f (s)$.       
Fig. 1(a) presents the evolution of $\bar f (s)$ versus time $s$ during 
some time period. This plot shows
that $\bar f (s)$ varies slightly between two successive time steps but
tends to increase in the long evolution process. 

Introduce another quantity, $F(s)$, the gap of the average fitness. The
definition of $F(s)$ is given as follows: Initial value of $F(s)$ is equal to
$\bar f (0)$. After $s$ updates, a large $F(s) > F(0)$ opens up. The current
gap F(s) is the maximum of all $F(s^{\prime})$, for all $0 \leq s^{\prime}
< s$. Fig. 1(b) shows $F(s)$ as a step-wising increasing function of $s$ during
the transient.
Actually, the gap is an envelope function that tracks the increasing peaks
in $\bar f (s)$. Indeed, punctuated equilibrium behavior appears in terms of
$\bar f (s)$.   

By definition [14], the separate instances when the gap $F(s)$ jumps to its
next higher value are separated by avalanches. Avalanches correspond to
plateaus in $F(s)$ during which $\bar {f}(s) < F(s)$. A new avalanche 
is initiated each time the gap jumps and ends up
when the gap jumps again. As the gap increases, the probability for the
average fitness, $\bar f (s)$, to fall below the gap increases also, and larger
and larger avalanches typically occur.

We can derive an exact gap equation of $F(s)$, similar to that found
in Ref. [16]. Suppose in the system the current gap 
is $F(s)$. If $F(s)$ is to be increased by $\Delta F$, i.e., from $F(s)$ to
$F(s)+\Delta F$, the average number of avalanches needed is $N_{\rm
av}=\Delta F L^d /(\bar f_c-F(s))$, where $\bar f_c$ is the critical value
of $\bar f (s)$. We can guarantee $N_{\rm av} \gg 1$ by
selecting $\Delta F \gg L^{-d}$. In the large $L$ limit, $N_{\rm av}$ can be
arbitrarily large. Hence, in this limit, the average number of time steps
required to increase the gap from $F(s)$ to $F(s)+\Delta F$ is given by the
interval $\Delta s=\langle S \rangle_{\rm F(s)} N_{\rm av}=\langle S
\rangle_{\rm F(s)} \Delta F L^d /(\bar f_c-F(s))$, where 
$\langle S \rangle _{\rm F(s)}$ 
is the average size of avalanche of the plateaus in the gap function.
>From the law of large numbers the fluctuation of this interval around its
average value vanishes. In the $\Delta F \rightarrow 0$ limit, $\Delta s
\rightarrow 0$. Taking the continuum limit we can obtain the differential
equation for $F(s)$,
\begin{equation}
\frac {\rm dF(s)} {\rm ds}=\frac {\bar f_c-{\rm F(s)}} 
{L^d \langle S \rangle _{\rm F(s)}}. 
\end{equation}

\noindent
Note this equation is exact. 

All SOC models, e.g., the BTW sandpile model [17], the earthquake models
[18], or Bak-Sneppen model [14], exhibit self-organized criticality in terms
of a power-law distribution of avalanche. It is natural to expect that we
can observe SOC in terms of the hierarchical structure of $\bar f(s)$, which
itself manifests complexity. 
It is simply another way to observe the same phenomenon 
by using such new quantity to define the avalanche,
which can be observed in different ways. 
As known, the emergence of complexity is independent of the
tools used to observe them provided that these tools are efficient and
strong enough. Similar to those used in Refs. [13,19], we present the
definition of $\bar {f}_0$-avalanche, where $\bar {f}_0$ 
($0.5 < \bar {f}_0 < 1.0$) is only a parameter 
used to define the avalanche. Suppose at time step $s_1$, 
$\bar {f}(s_1)$ is larger than $\bar {f}_0$. If , at time step $s_1$+1, $\bar
{f}(s_1+1)$ is less than $\bar {f}_0$, this initiates a creation-annihilation
branching process. The avalanche still continues at time step $s^{\prime}$, 
if all the $\bar {f}(s)$ are less than $\bar {f}_0$ for $1 \leq s \leq 
s^{\prime}-1$. And the avalanche stops, say, at time step $s_1+S$, 
when $\bar {f}
(s_1+S)>\bar {f}(s_1)$. In terms of this definition, the size of the
avalanche is the number of time steps between subsequent punctuation of the
barrier $\bar {f}_0$ by the signal $\bar {f} (s)$. In the above example, the
size of the avalanche is $S$. It can be clearly seen from Fig. 1(a) that this
definition guarantees the hierarchical structure of avalanches-- larger
avalanches consists of smaller avalanches. As $\bar f_0$ is lowered, bigger
avalanches are subdivided into smaller ones. Hence, the statistics of $\bar
f_0$-avalanche will inevitably have a cutoff if $\bar f_0$ is not chosen to
be $\bar f_c$. We can
also define  $\bar f_c$-avalanche. Nevertheless, $\bar f_0$-avalanche in
the stationary state has the same scaling behavior as $\bar f_c$-avalanche
provided $\bar f_0$ close to $\bar f_c$. We measure $\bar
f_0$-avalanche distribution for one-dimensional (1D)
and two-dimensional (2D) Bak-Sneppen models.
The simulation results are given in Fig. 2. The exponent $\tau$, defined by
$P(S) \sim S^{- \tau}$, is 1.800 for 1D model and 1.725 for 2D model. Another
exponent, $D$, avalanche dimension [13], defined by $n_{\rm cov} \sim
S^{D/d}$, where $n_{\rm cov}$ is the number of sites covered by an
avalanche, and $d$ is the space dimension, is measured. We find $D$=2.45
for 1D model and 3.10 for 2D model.          
  
Up to now, a question is still unsolved. It is  the critical value of
$\bar f$, $\bar f_c$. This may be troublesome if the system size is finite,
but when we consider large $L$ limit, it can be
easily accomplished. Recall the evolution of Bak-Sneppen model, or the
detailed research of the  model of Ref. [13], the densities of sites 
with random
numbers is uniform above $G$ and vanishes below $G$ when $L \rightarrow
\infty$, where $G$ is the gap of extremal site.
One can readily obtain 
\begin{equation}
\lim_{L \rightarrow \infty}\bar f (s)=
\lim_{L \rightarrow \infty}\frac {1+G(s)} {2} .
\end{equation}

\noindent
From Eq. (3) one can immediately obtain
\begin{equation}
\lim_{L \rightarrow \infty} \bar {f}_c=\lim_{L \rightarrow \infty}
\frac {1+f_c} {2} .
\end{equation}

\noindent
Hence, $\bar {f}_c$ can be easily determined from Eq. (4). Using the results
of $f_c$ provided by Refs. [13,20], one can obtain $\bar {f}_c$, 0.83351 for
1D model and 0.66443 for 2D model. However, Eqs. (3) and (4) are not valid
for a finite-size system, since one can not ensure the distribution of
random numbers during a
finite-size system is really uniform. Due to the fluctuation of $\bar {f}(s)$ 
it is extremely difficult to determine the exact critical value of $\bar {f}$ 
for a finite-size system. One may estimate $\bar {f}_c$ for a finite-size
system using the simulation. We find that this value weakly depends on the 
system size when it is enough large and $\bar f_c$ will approach the
corresponding value for infinite systems. Fig. 3 shows the fluctuation 
of $\bar f$ for a
1D model of size $L=200$ near the critical state. We note, in
this curve, $\bar f$ fluctuates slightly around some average value and 
does not tend
to increase any more during a long time period. We may say 
that the system approaches
its stationary state. In this sense, we suggest that $\bar f$ may be a good
quantity in determining the emergence of criticality. That is, the great
fluctuation of $f_{\rm min}$ will not affect us to determine when we approach
the critical state. We need only to know the feature of $\bar f$. It is more
reasonable and easily accepted since $\bar f$ is a global quantity and
condenses information of the system and its components.   
    
Why we call the $\bar f_0$-avalanche a different hierarchy of avalanches? 
Firstly,
this new type of avalanche is defined on the global level, in terms of the new
global quantity, $\bar f$. The background of this definition is different
from those used before. This type of avalanche reflects the fractal
geometry in terms of the global feature. Secondly, one can notice that the
exponents $\tau$ of avalanche distribution obtained in our simulations 
are different from those found in Ref. [13].
From this point of view, one can conclude that this type of avalanche is
different from any one observed before.

SOC was suggested by Bak et al. to be the "fingerprints"
of a large variety of complex system, which is represented by a scale-free 
line on a double logarithm plot. In order
to know the criticality of a system one needs to know when the system
reaches the stable stationary state where the phase transition occurs. It is
extremely difficult and nearly impossible for one to know when a system 
in nature
approaches its critical state. One has to study the
ubiquitous fractal geometrical structure carved by avalanches through
thousands of millions of years. However, in laboratory experiments and
computer simulations, one needs a criterion to determine whether stationary state
approaches, even, reaches, since statistics of avalanches may only be done
under critical state of the system. Given Bak-Sneppen model, when the
extremal signal, $f_{\rm min}$, approaches the self-organized threshold,
$f_c$, the ecosystem reaches its stationary state. However, $f_{\rm min}$
itself fluctuates greatly time to time, which brings a big problem in
determining the appearance of criticality. Thus, we provide a new quantity,
$\bar f$, for a candidate in judging the emergence of criticality. As shown,
$\bar f$ is relatively stable in a short time period. Hence, when $\bar f$
does not tend to increase any more, one may say that the system approaches
its stationary state. And, one can observe criticality in a quite long time
period.

In conclusion, a different hierarchy of avalanches is observed in Bak-Sneppen
model. A new quantity, $\bar f$, is presented and suggested by us to be a
possible candidate in determining the emergence of criticality. An exact gap
equation and simulation results are also given.

This work was supported in part by NSFC in China and Hubei-NSF. 
We thank Prof. T. Meng for correspondence and helpful discussions.
X.C. thanks J. Schukraft and D. Jouan for hospitalities during his visits in
CERN and Orsay separately.

\vskip 0.5cm
\begin{center}
\bf {Figure Captions}
\end{center}

\vskip 0.2cm
Fig. 1: (a) The variation of $\bar f$ versus time during a time period for a
            one-dimensional Bak-Sneppen model of size $L=200$. 
	    This shows the hierarchical structure of $\bar f$. 
	(b) Punctuated equilibrium of $\bar f$ for a one-dimensional
	    Bak-Sneppen model of size $L=200$. We track the increasing 
	    signal of $\bar f(s)$, i.e., F(s). 

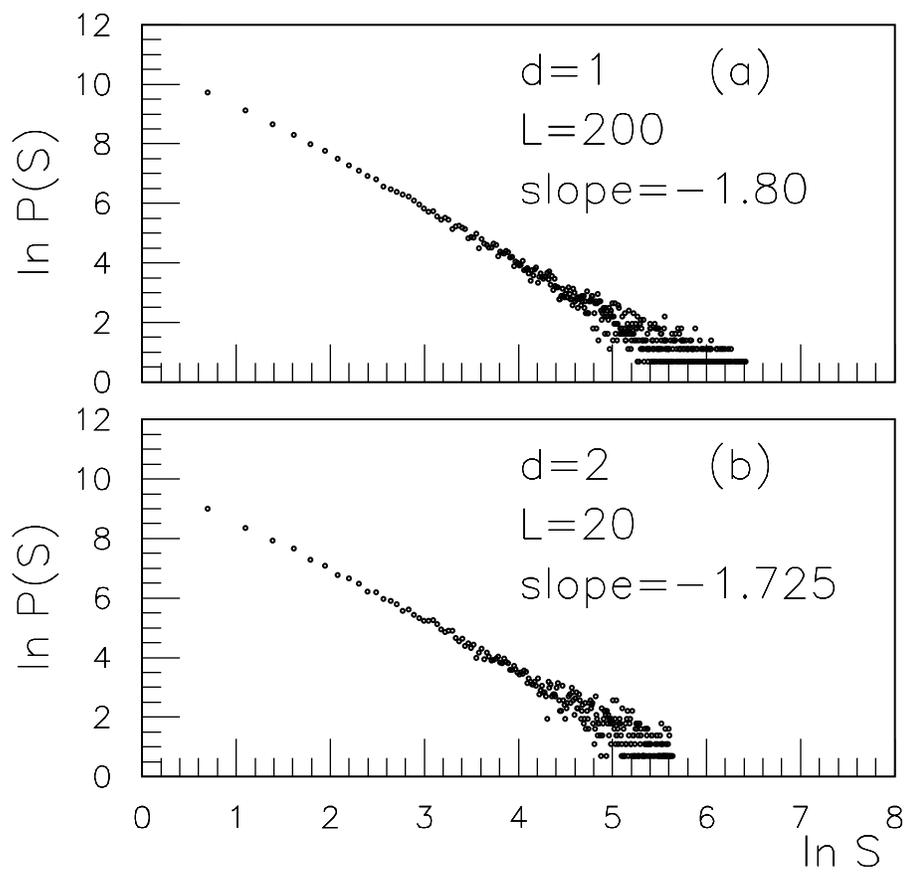
Fig. 2: Distributions of $\bar f_0$-avalanche for  
        (a) one-dimensional Bak-Sneppen model with size $L=200$,
	$\bar{f_0}=0.821$ and slope=-1.800; and 
	(b) two-dimensional Bak-Sneppen model with size $L=20$,
        $\bar{f_0}=0.648$ and slope=-1.725.

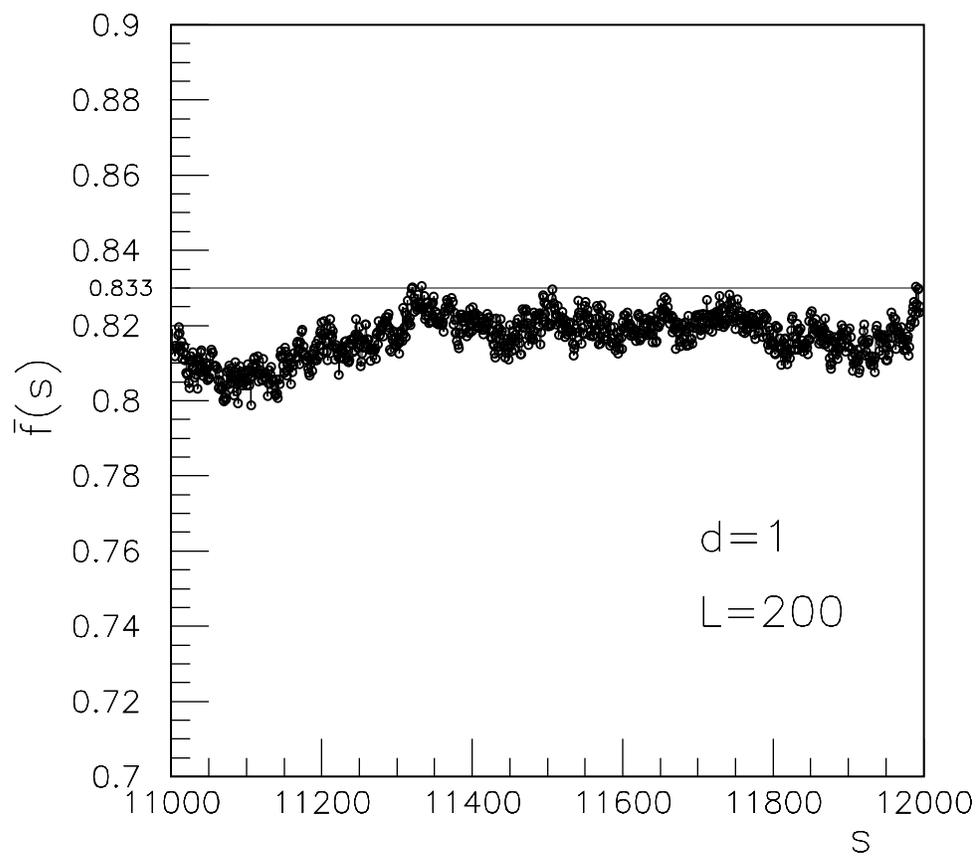
Fig. 3: The fluctuation of $\bar f$ around the critical state for a
	one-dimensional Bak-Sneppen model of size $L=200$.
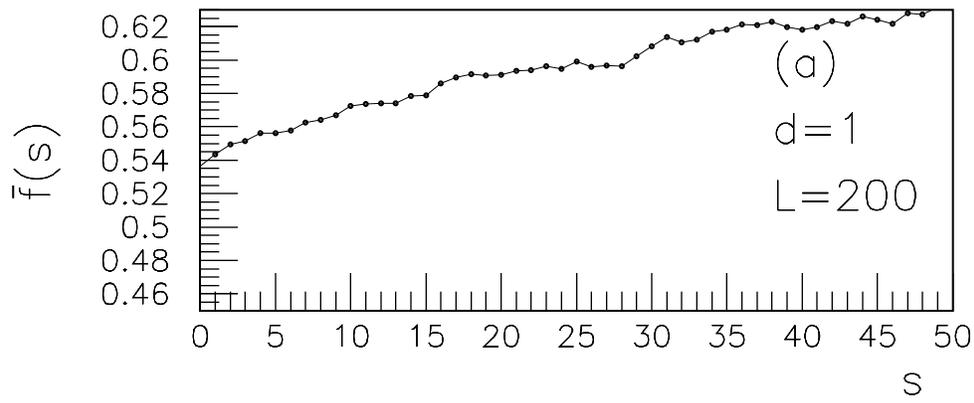
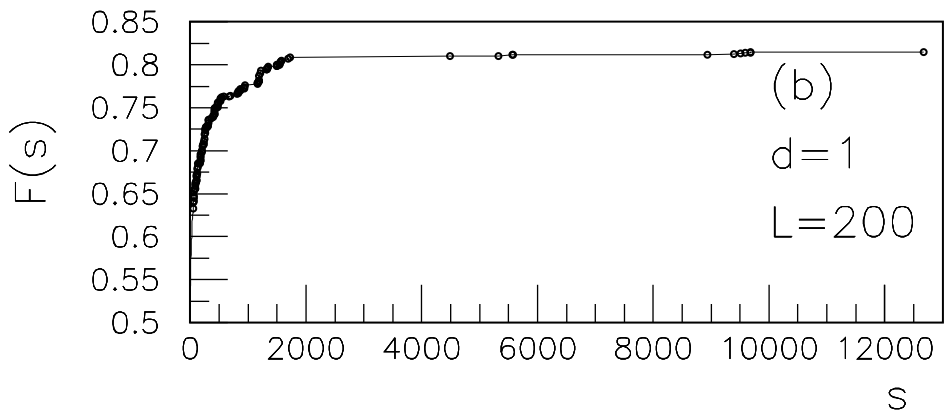
\end{document}